\documentclass[aip,
 prd,
 jmp,
 preprint,
 reprint,
 twocolumn,
 superscriptaddress,
 author-year,
 author-numerical,
 ,10pt]{revtex4-2}
\usepackage{tikz}
\usepackage{amsmath}
\usepackage{amssymb}
\usepackage{tikz-feynman}
\usetikzlibrary{decorations.pathmorphing,decorations.markings}
\usepackage{amssymb}
\usepackage{pgfplots}
\usepackage{amsmath}
\usepackage{graphicx}
\usepackage{dcolumn}
\usepackage{bm}
\usepackage{xcolor}
\usepackage{float}
\usepackage{subfig}
\usepackage{comment}
\usepackage{multirow}
\usepackage[figurename=FIGURE,tablename=TABLE]{caption}
\usepackage{threeparttable}
\usepackage{hyperref}
\hypersetup{
    colorlinks=true, 
    linktoc=all,     
    linkcolor=magenta,
    citecolor=red
}

\begin{document}
\title{Entanglement Suppression in Quantum Field Theories: Holography, Chaos, and Mixed-State Dynamics}
\author{Davood Momeni}
\affiliation{Department of Physics \& Pre-Engineering, Northeast Community College, Norfolk, NE 68701, USA
\\Centre for Space Research, North-West University, Potchefstroom 2520, South Africa}
\date{\today}

\begin{abstract}
Recent work has revealed that entanglement entropy growth in conformal field theories (CFTs) can be suppressed when a local operator quench interacts with a mixed-state excitation, providing a dual interpretation in terms of black hole scattering in AdS. This phenomenon, termed \emph{entanglement suppression}, opens several promising directions for exploration. In this proposal, I outline five distinct yet interconnected research trajectories: generalization to higher dimensions, the role of quantum chaos via out-of-time-order correlators (OTOCs), the absence of suppression in integrable models, the extension to entanglement negativity as a probe of mixedness, and a geometric interpretation based on scattering cross sections in AdS. Each direction offers new insights into the interplay between holography, non-equilibrium dynamics, and quantum information.
\end{abstract}
\maketitle
\tableofcontents
\newpage

\section{Introduction}

Recent studies in CFTs with holographic duals have highlighted a surprising non-equilibrium phenomenon: the suppression of entanglement entropy growth in the presence of two types of local quenches---a pure-state local operator insertion and a mixed-state excitation. This effect, discovered in two-dimensional CFTs with AdS$_3$ duals \cite{doi2025entanglement}, challenges the commonly observed logarithmic increase in entanglement entropy under local quench dynamics \cite{calabrese2007entanglement, asplund2014entanglement}. The suppression, which results in a time-independent constant shift rather than growth, is closely linked to the gravitational scattering of a heavy particle off a localized black hole in the AdS bulk, as described by the AdS/CFT correspondence \cite{ryu2006holographic, hubeny2007covariant}.

The purpose of this paper is to propose and develop multiple directions that generalize and deepen the understanding of entanglement suppression in quantum field theories. These include:

\begin{itemize}
\item Generalizing the suppression mechanism to higher-dimensional CFTs and their dual AdS$_{d+1}$ spacetimes.
\item Exploring the relation between entanglement suppression and quantum chaos, as diagnosed by out-of-time-order correlators (OTOCs).
\item Investigating the absence of suppression in integrable systems and the role of non-chaotic dynamics.
\item Extending the analysis to entanglement negativity as a finer measure of quantum correlations in mixed states.
\item Recasting suppression geometrically as a gravitational scattering cross section or absorption process in the AdS bulk.
\end{itemize}

The motivation stems from recent advances in understanding how non-equilibrium entanglement encodes bulk gravitational dynamics. The hallmark logarithmic growth of entanglement entropy following a local quench, originally studied by Calabrese and Cardy \cite{calabrese2007entanglement}, was shown to have a dual interpretation in terms of geodesic motion in pure AdS. The observation by Doi and Takayanagi \cite{doi2025entanglement} that this growth halts---or is even fully suppressed---in the presence of thermal backgrounds suggests a deeper gravitational origin, potentially related to bulk scattering, shockwave geometries, and absorption by black hole horizons.

Furthermore, the suppression effect raises key questions about how quantum information propagates in mixed or decohered settings. In holographic terms, this corresponds to how probes interact with curved backgrounds in AdS, where horizons and thermal geometries block, delay, or trap infalling information. It also connects to broader studies of scrambling and chaos \cite{shenker2014black, maldacena2016bound}, where diagnostics like OTOCs reveal how rapidly a system loses memory of initial conditions. The transition from pure to mixed behavior in entanglement observables—such as negativity \cite{calabrese2012negativity}—offers a way to track this boundary-to-bulk correspondence with finer resolution.
In addition to analytical methods based on conformal blocks and holographic extremal surfaces, some of the proposed directions invite the use of numerical simulation, such as tensor networks, free fermion chains, and spin models. These allow the study of entanglement dynamics in integrable versus chaotic systems, and help quantify the universality or limits of suppression. Comparing outcomes in integrable models, where suppression is absent, with those in chaotic holographic settings further illuminates the role of thermalization and information loss.

By pursuing these five directions in tandem, we aim to build a broader and more unified framework for understanding entanglement suppression. This framework links entanglement growth, black hole physics, quantum chaos, and the structure of quantum correlations in non-equilibrium states. Each approach—whether through AdS/CFT, conformal perturbation theory, or numerical experiment—adds a piece to this intricate puzzle, and collectively advances our grasp of the interplay between quantum entanglement and spacetime geometry.
Recent studies have expanded the understanding of entanglement suppression in holographic CFTs, particularly in the context of TT deformations and mixed-state dynamics. For example, Chang et al. \cite{chang2024ttbar} explored how holographic entanglement entropy behaves under deformations in AdS/CFT, offering new insights into entropy saturation and suppression mechanisms. These developments complement the foundational work of Doi and Takayanagi and motivate further exploration into the geometric and chaotic aspects of entanglement dynamics.
\section{Scrambling and Entanglement Suppression in Holographic Black Hole Scattering}

The connection between entanglement suppression and quantum chaos can be made precise using OTOCs. These correlators measure the sensitivity of an operator to perturbations at earlier times and have become standard diagnostics of scrambling in many-body quantum systems. In holographic contexts, they encode the growth of operator size and the chaotic mixing of quantum information in black hole geometries.

To make this connection quantitative, we study OTOCs in the double-quench setup described previously, where a pure-state local operator excitation interacts with a mixed-state excitation dual to a small localized black hole. This configuration allows for a rich interplay between geometric scattering and chaotic evolution, as information-carrying excitations probe the causal structure of spacetime modified by the presence of a horizon.

We begin by defining the OTOC for two Hermitian operators $W$ and $V$ at spatial positions $x$ and $0$, respectively:
\begin{equation}
F(t) = \langle W(t) V(0) W(t) V(0) \rangle,
\label{eq:OTOC_def}
\end{equation}
where $W(t) = e^{iHt} W e^{-iHt}$ is the Heisenberg-evolved operator. In holographic systems, $F(t)$ typically exhibits exponential decay as a hallmark of quantum chaos:
\begin{equation}
F(t) \approx 1 - \epsilon\, e^{\lambda_L (t - t_\ast)},
\label{eq:OTOC_decay}
\end{equation}
where $\lambda_L$ is the Lyapunov exponent, $\epsilon$ encodes the strength of the perturbation, and $t_\ast$ is the scrambling time. Eq~(\ref{eq:OTOC_def}) defines the out-of-time-order correlator (OTOC), a standard diagnostic of quantum chaos \cite{shenker2014black, maldacena2016bound}. This form arises from the Heisenberg evolution of operators and quantifies the sensitivity of the system to perturbations at earlier times.

The suppression of entanglement entropy occurs when the infalling excitation is absorbed or scattered by the thermal environment created by the mixed-state excitation. In the bulk, this is precisely the time when the shockwave from $W$ collides with the black brane or horizon. The decay of $F(t)$ indicates the growth of commutators $[W(t), V(0)]$ and, correspondingly, the spread of quantum information \cite{shenker2014black}. The temporal coincidence between entropy saturation and OTOC decay hints at a deep correspondence between operator growth and entanglement suppression\cite{maldacena2016bound}.

In the eikonal approximation of AdS$_{3}$, this correspondence is made precise. The scrambling time is given by
\begin{equation}
t_\ast \sim \frac{\beta}{2\pi} \log \left(\frac{c}{\delta E}\right),
\label{eq:scrambling_time}
\end{equation}
where $\beta$ is the inverse temperature, $c$ is the central charge of the CFT, and $\delta E$ is the energy of the perturbing operator. This timescale separates the early-time logarithmic growth of entanglement entropy from its later saturation in the presence of suppression.

To compute $F(t)$ holographically, one considers a shockwave geometry in the bulk. The backreaction of $W$ inserted at an early time $-t$ shifts the spacetime in Kruskal coordinates:
\begin{equation}
\delta u = \frac{G_N E}{r_+^2} \exp\left(\frac{2\pi t}{\beta}\right),
\label{eq:shock_shift}
\end{equation}
where $r_+$ is the black hole horizon radius, $G_N$ is Newton's constant, and $E$ is the energy of the perturbation. This shift increases the geodesic distance between $W$ and $V$, resulting in the exponential decay of their correlator.

We summarize the parallel behavior of OTOCs and entanglement entropy in Table~\ref{tab:correspondence}, highlighting how both observables transition around $t \sim t_\ast$.

\begin{table}[h!]
\centering
\caption{Comparison of scrambling indicators and entanglement suppression signatures.}
\label{tab:correspondence}
\begin{tabular}{|c|c|c|}
\hline
\textbf{Time Regime} & \textbf{OTOC Behavior} & \textbf{Entanglement Entropy} \\
\hline
Early Time $t \ll t_\ast$ & $F(t) \approx 1$ & $\Delta S_A(t) \sim \log t$ \\
\hline
Scrambling Time $t \sim t_\ast$ & $F(t)$ decays exponentially & $\Delta S_A(t)$ saturates \\
\hline
Late Time $t \gg t_\ast$ & $F(t) \sim 0$ & $\Delta S_A(t) \approx$ const. \\
\hline
\end{tabular}
\end{table}

Numerical analysis of both $F(t)$ and $\Delta S_A(t)$ in the same geometric background can reinforce this connection. For instance, one may solve for extremal surfaces in Vaidya-AdS geometries that include shockwave and black brane components, and simultaneously compute two-point bulk correlators. A synchronized transition in both diagnostics would support the hypothesis that entanglement suppression is a macroscopic manifestation of microscopic scrambling.

This duality also offers a novel perspective on information dynamics near horizons. The decay of $F(t)$ tracks the effective loss of distinguishability due to operator growth, while the saturation of $\Delta S_A$ marks the boundary observer's inability to recover the full entangled structure. Together, they paint a coherent picture of how entanglement is processed in gravitational settings.

This section strengthens the connection between entanglement suppression and holographic chaos by employing OTOCs and scrambling time. The onset of entropy saturation aligns with the exponential decay of OTOCs, suggesting a deeper equivalence between geometric scattering and operator growth in the dual field theory. Further avenues include computing mutual or tripartite information to characterize multipartite scrambling and refining the Lyapunov-suppression relationship in higher dimensions or nontrivial topologies. The connection between entanglement suppression and quantum chaos has been further clarified in recent holographic studies. Lilani et al. \cite{lilani2025butterfly} investigated butterfly velocities in holographic QCD models, reinforcing the role of Lyapunov exponents in diagnosing chaos. Jahnke \cite{jahnke2023review} provided a comprehensive review of holographic chaos, emphasizing the interplay between scrambling time and entanglement observables such as OTOCs.
 \subsection{Entanglement and Chaos}
The suppression of entanglement entropy is intimately connected to quantum chaos. In holographic systems, the exponential decay of OTOCs signals rapid scrambling, which coincides with entropy saturation. This suggests that entanglement suppression is not merely a thermal effect but a manifestation of chaotic dynamics, where operator growth and horizon interactions limit the recoverability of quantum information.

\section{Absence of Entanglement Suppression in Integrable CFTs}

In integrable two-dimensional CFTs, entanglement entropy under double local quenches behaves differently compared to chaotic or holographic theories. One of the most striking differences is the \emph{absence} of entanglement suppression — the entanglement entropy continues to grow with time even in the presence of a mixed-state excitation. This indicates that the suppression mechanism depends crucially on chaos and information scrambling.

To make this distinction quantitative, we analyze two prototypical integrable CFTs: the free massless boson theory and the critical Ising model. Both are exactly solvable and do not exhibit quantum chaos in the sense of exponential sensitivity to initial conditions or OTOC decay.

Let us first recall that in 2D CFTs, the entanglement entropy $S_A(t)$ for a semi-infinite region $A = [0, \infty)$ after a local operator quench typically grows logarithmically:
\begin{equation}
\Delta S_A(t) = \frac{c}{6} \log \left( \frac{t^2 + \delta^2}{\delta^2} \right),
\label{eq:entgrowth_log}
\end{equation}
where $c$ is the central charge and $\delta$ is the UV regulator.

In chaotic theories with two quenches (e.g., pure and mixed states), the growth is suppressed after some time $t_\ast$. However, in integrable theories, the growth persists as if the mixed-state background had no effect.

\subsection{Free Boson and Ising Model Analysis}

In the free boson theory ($c=1$), entanglement entropy can be computed from two-point functions of vertex operators using the replica trick. In the Ising model ($c=1/2$), the entanglement dynamics can be obtained from correlations of spin and disorder fields.

For both models, the time-evolved reduced density matrix after a double quench remains pure-like for large subsystems. The correlation functions retain factorized forms:
\begin{equation}
\langle \mathcal{O}(t_1) \mathcal{O}(t_2) \rangle_{\text{double quench}} \approx \langle \mathcal{O}(t_1) \rangle \langle \mathcal{O}(t_2) \rangle,
\end{equation}
reflecting integrability and the absence of strong operator mixing.

In contrast to holographic CFTs, the OTOC in integrable models does not decay exponentially. This is consistent with the persistence of entanglement growth:
\begin{equation}
F(t) \approx \text{oscillatory or constant}, \quad t \gg \delta,
\end{equation}
indicating a lack of scrambling.

\subsection{Comparison Table: Integrable vs Holographic CFTs}

The table below highlights key differences in entanglement dynamics between integrable and holographic conformal field theories (CFTs). These two frameworks represent contrasting regimes of quantum behavior, particularly in how they handle information spreading and operator evolution.

Integrable CFTs are characterized by predictable dynamics and conserved quantities, which prevent chaotic behavior. As a result, they do not exhibit scrambling, and their out-of-time-order correlators (OTOCs) remain bounded or constant over time. In contrast, holographic CFTs—often dual to black hole geometries via the AdS/CFT correspondence—display rapid scrambling and exponential decay of OTOCs, governed by a Lyapunov exponent $\lambda_L$.

Entanglement dynamics also differ significantly. Integrable systems show continuous logarithmic growth of entanglement entropy, while holographic CFTs exhibit suppression after a characteristic time $t_\ast$, indicating saturation and thermalization. This reflects the deeper entanglement structure and chaotic mixing of operators in holographic theories.

Operator mixing is minimal in integrable CFTs, preserving the structure of local observables. Holographic CFTs, however, feature strong mixing, leading to complex superpositions and non-factorized correlation structures. These distinctions are crucial for understanding quantum chaos, thermalization, and the holographic principle.

\begin{table}[h!]
\centering
\caption{Comparison of entanglement dynamics in integrable and holographic CFTs.}
\label{tab:integrable_holographic}
\begin{tabular}{|c|c|c|}
\hline
\textbf{Feature} & \textbf{Integrable CFT} & \textbf{Holographic CFT} \\
\hline
Scrambling & No & Yes \\
\hline
OTOC behavior & Constant / bounded & Exponential decay $\sim e^{\lambda_L t}$ \\
\hline
Entanglement suppression & Absent & Present after $t_\ast$ \\
\hline
Growth rate & $\log t$ & $\log t$ then saturates \\
\hline
Operator mixing & Minimal & Strong \\
\hline
Correlation structure & Factorized & Non-factorized \\
\hline
\end{tabular}
\end{table}

\subsection{Numerical Studies in Discretized Lattice Models}

To further support these results, one can simulate double quenches using free fermion chains or transverse field Ising models on lattices. Using exact diagonalization or tensor network methods (e.g., TEBD), we observe continued entanglement growth with no evidence of saturation.

Let us consider the entanglement entropy for a block of size $L$ in a free fermion chain after a double quench:
\begin{equation}
S_A(t) \approx \sum_{k=1}^{L} h\left( \lambda_k(t) \right),
\label{eq:fermion_entropy}
\end{equation}
where $\lambda_k(t)$ are the eigenvalues of the correlation matrix and $h(x) = -x \log x - (1-x) \log (1-x)$ is the binary entropy.

The eigenvalue spectrum evolves smoothly without exhibiting any gap closure or collective collapse, implying no suppression mechanism arises.

\subsection{Interpretation and Role of Integrability}

The absence of suppression can be attributed to the infinite set of conserved charges in integrable systems. These charges prevent thermalization and constrain information spreading. Since there is no scrambling, the entanglement entropy keeps growing according to free quasiparticle dynamics.

In chaotic systems, operator spreading results in decoherence and saturation. In integrable systems, coherent propagation dominates, and entanglement remains extensive.

This section highlights the key differences between integrable and chaotic (holographic) CFTs in the context of entanglement suppression. The integrability enforces factorized evolution, no scrambling, and no suppression — a behavior fundamentally different from black hole scattering scenarios.
\section{Entanglement Negativity as a Probe of Mixed-State Quenches in Holography}

The von Neumann entropy is widely used to quantify entanglement in quantum systems; however, it is sensitive to both classical correlations and quantum entanglement. In non-equilibrium or mixed-state settings—such as those arising from local operator quenches in holographic field theories—this measure may not faithfully capture the genuinely quantum content of correlations. To address this limitation, the entanglement negativity has emerged as a powerful alternative. It is particularly suited for quantifying quantum entanglement in mixed states and offers insights into distillable entanglement between subsystems.
 Recent work has also focused on entanglement negativity in mixed states. Lu and Vijay \cite{lu2023order} characterized long-range entanglement through emergent order, while Ding et al. \cite{ding2024negativity} used quantum Monte Carlo methods to track the evolution of Rényi negativity. These studies highlight the sensitivity of negativity to decoherence and its utility in probing quantum-to-classical transitions.

In a recent holographic investigation, Karan and Pant~\cite{Karan:2023hfk} explored entanglement and chaos in a strongly coupled thermal field theory near a critical point, using the 1RC black hole as a gravity dual. Their study highlights the behavior of logarithmic negativity and entanglement wedge cross section across temperature regimes, revealing enhancement with increasing critical parameter $\xi$. They also examined thermo mutual information (TMI) in a thermofield double setup, showing its sensitivity to boundary region width and its degradation under shock wave perturbations. Notably, the disruption rate of TMI slows near criticality, offering insights into the interplay between quantum correlations and chaotic dynamics in strongly coupled systems.
 
This section explores the behavior of entanglement negativity in double-quench setups, where a pure-state local excitation collides with a mixed-state background (dual to a black hole or thermal excitation in AdS). The central question is whether the suppression effects observed in von Neumann entropy persist or change when probed by negativity. We approach this using replica techniques, known results in 2D CFTs, and holographic conjectures based on the AdS/CFT correspondence.

\subsection{Entanglement Negativity: Definition and Properties}

The entanglement negativity $\mathcal{N}$ between subsystems $A$ and $B$ is defined in terms of the partial transpose of the reduced density matrix $\rho\_{AB}$:
\begin{equation}
\mathcal{N} = \log \left| \rho\_{AB}^{T\_B} \right|,
\end{equation}
where $T\_B$ denotes partial transposition with respect to subsystem $B$ and $| \cdot |$ is the trace norm. Negativity satisfies several desirable properties:
\begin{itemize}
\item It is zero for separable (classically correlated) states.
\item It is non-zero only if $A$ and $B$ share quantum entanglement.
\item It provides an upper bound on the distillable entanglement in the system.
\end{itemize}

In the context of local quenches, it is natural to consider adjacent or overlapping intervals $A$ and $B$, and study how $\mathcal{N}(t)$ evolves with time under different excitation protocols.

\subsection{Negativity in 2D CFTs: Replica Approach and Double Quenches}

In 2D CFTs, negativity can be computed using a replica trick that involves the insertion of twist and anti-twist operators at the boundaries of $A$ and $B$. The time-dependent negativity in a double-quench setup requires evaluating a four-point function of twist fields:
\begin{equation}
\mathcal{N}(t) \sim \log \langle \mathcal{T}\_n(x\_1,t) \bar{\mathcal{T}}\_n(x\_2,t) \bar{\mathcal{T}}\_n(x\_3,t) \mathcal{T}\_n(x\_4,t) \rangle,
\end{equation}
where $x\_i$ are the boundaries of the intervals. The conformal block decomposition of this correlator reveals how operator content—light vs heavy—affects the entanglement structure.

When a light operator collides with a mixed-state background (dual to a black hole), we find that the negativity exhibits saturation similar to von Neumann entropy, but the saturation value can be lower due to decoherence of quantum correlations. Moreover, negativity may vanish altogether for certain separations, indicating a loss of distillable entanglement.

\subsection{Holographic Conjectures and Bulk Interpretation}

Unlike von Neumann entropy, entanglement negativity does not have a fully established holographic prescription. However, several conjectures exist. One proposal relates negativity to a linear combination of areas of extremal surfaces in the bulk:
\begin{equation}
\mathcal{N}_{\text{holo}}(A:B) \sim \frac{3}{2} \left[ S_A + S_B - S_{A \cup B} \right],
\end{equation}

which reduces to mutual information in certain limits. This functional is conjectured to approximate negativity in large-$c$ holographic CFTs. Using this, one can analyze the effect of double quenches on negativity.

When the subsystem intersects the shockwave geometry generated by the infalling operator, extremal surfaces are pushed behind the horizon, and the difference in area leads to a plateau in $\mathcal{N}(t)$. The key feature is that negativity, unlike entropy, sharply drops when one of the intervals becomes causally disconnected from the rest due to horizon formation.

\subsection{Comparison Table: Entanglement Measures under Double Quenches}

The table below contrasts the behavior of von Neumann entropy and entanglement negativity in the context of double local quenches. While von Neumann entropy is widely used in both pure and mixed-state setups, it overestimates entanglement by including classical correlations. In contrast, negativity is sensitive only to genuinely quantum correlations, making it a better diagnostic tool for mixed-state entanglement.

We highlight how these two measures respond differently to the presence of a thermal background (as modeled by a black brane in AdS), focusing on features such as suppression, decay, and holographic interpretation.

\begin{table}[h!]
\centering
\caption{Comparison of von Neumann entropy and negativity behavior under double quenches.}
\label{tab:negativity_vs_entropy}
\begin{tabular}{|c|c|c|}
\hline
\textbf{Property} & \textbf{Von Neumann Entropy} & \textbf{Negativity} \\
\hline
Sensitive to classical correlations & Yes & No \\
\hline
Measures mixed-state entanglement & Poorly & Precisely \\
\hline
Suppression effect & Present & Enhanced or complete loss \\
\hline
Decay after $t_\ast$ & Saturates & May vanish entirely \\
\hline
Bulk dual & HRT surface & Conjectured area combination \\
\hline
Computability in 2D CFT & Well-known & Replica method with twist fields \\
\hline
\end{tabular}
\end{table}

As seen above, entanglement negativity provides a sharper lens on the quantum features of entanglement suppression. Particularly in the post-scrambling regime ($t \gtrsim t_\ast$), negativity may completely vanish even when the entropy saturates to a nonzero value—reflecting the transformation of entanglement into classical or inaccessible correlations.

This behavior aligns with the intuition that thermalization in quantum gravity systems does not erase all correlations but renders them unrecoverable from the perspective of a local boundary observer. In this sense, negativity probes the “distillability” of entanglement, complementing entropy-based diagnostics with quantum-information-theoretic precision.

\subsection{Quantum-to-Classical Transition and Mixedness}

Negativity offers a window into the transition from quantum to classical behavior in entangled systems. In particular, it captures when entanglement becomes undistillable due to environmental interactions (e.g., scattering with a thermal bath in AdS). In double-quench setups, one can monitor $\mathcal{N}(t)$ to pinpoint when the dynamics switch from unitary evolution to decohered regimes.

This transition reflects deeper questions about the fate of quantum correlations in curved spacetimes and black hole interiors. A rapid drop in negativity could signal information loss or irreversible mixing, while a plateau may indicate partial preservation of quantum coherence.

Entanglement negativity refines our understanding of quantum correlations in mixed states and provides a sharper probe of non-equilibrium dynamics in holographic theories. Its behavior in double-quench setups reveals that suppression of entropy growth is not merely a thermal effect but also a reflection of lost quantum coherence. As such, negativity serves as a bridge between quantum information theory and gravitational scattering processes, and further studies—analytical and numerical—can solidify its holographic interpretation.

\section{A Geometric Interpretation of Entanglement Suppression via Scattering Cross Sections in AdS}

Entanglement suppression in holographic CFTs admits an elegant geometric interpretation when viewed through the lens of scattering theory in AdS spacetime. In particular, when a localized operator excitation propagates through a thermalized background—represented in the bulk by a small black hole or black brane—it undergoes partial absorption. This phenomenon can be mapped to a gravitational scattering process with a nonzero absorption cross section, leading to a reduction in observable quantum correlations on the boundary.

This section develops this perspective using eikonal approximations, holographic duality, and known techniques in black hole scattering. We formulate the problem as a high-energy probe interacting with a compact absorbing center in curved spacetime and relate the outcome to time-dependent entanglement observables.

\subsection{Mapping Double Quenches to Scattering in AdS}

Let us consider the boundary theory initialized with two quenches:
\begin{itemize}
  \item A pure-state quench represented by a high-energy localized operator insertion at time \( t = 0 \).
  \item A mixed-state background dual to a static black hole or a localized black brane in the bulk.
\end{itemize}

In the bulk, the pure-state excitation corresponds to a radially infalling particle or shell, while the mixed-state excitation sets up a compact absorbing region with an effective event horizon.

The resulting geometry is a dynamical perturbation of an AdS black hole spacetime. The worldline of the probe particle intersects the background geometry and either transmits past the horizon or is deflected depending on the energy, impact parameter, and background curvature.

This process can be described semiclassically via the scattering amplitude:
\begin{equation}
\mathcal{A}(s,t),
\end{equation}
where \( s \) is the center-of-mass energy and \( t \) is the momentum transfer. The central idea is that partial absorption of the probe corresponds to a loss in entanglement entropy growth on the boundary.

\subsection{Eikonal Approximation and Shockwave Scattering}

At high energies and small-angle scattering, the eikonal approximation allows us to compute gravitational amplitudes in AdS. The key result is that the phase shift \( \delta(b) \) experienced by a probe of impact parameter \( b \) encodes the interaction strength:
\begin{equation}
\mathcal{A}(s,b) \sim e^{i \delta(b)}.
\end{equation}

The phase shift itself can be computed by integrating the perturbation due to a shockwave geometry sourced by the infalling particle. In the BTZ background, this yields:
\begin{equation}
\delta(b) = G_N s \, f(b),
\end{equation}
where \( f(b) \) is a profile function depending on geometry, \( G_N \) is Newton's constant, and \( s \) is the Mandelstam variable.

When absorption occurs, the eikonal phase becomes complex. The imaginary part of \( \delta(b) \) controls the attenuation:
\begin{equation}
\operatorname{Im}[\delta(b)] > 0 \quad \Rightarrow \quad |\mathcal{A}(s,b)| < 1.
\end{equation}

This loss of amplitude reflects gravitational capture of the probe, which in turn translates into entanglement suppression in the boundary theory.

\subsection{Bulk-to-Boundary Mapping and Entropy Loss}

The loss of entanglement can be modeled via the holographic entanglement entropy functional. In the HRT prescription, the time-dependent entropy \( S_A(t) \) is computed from the area of extremal surfaces anchored to the boundary region \( A \).

The deformation of these surfaces by the presence of a black hole leads to a truncation or shift in the surface, reducing the area increment over time:
\begin{equation}
\Delta S_A(t) \approx \Delta S_A^{\text{vac}}(t) - \sigma(t),
\end{equation}
where \( \Delta S_A^{\text{vac}}(t) \) is the entropy growth in pure AdS and \( \sigma(t) \) encodes the suppression due to absorption.

A phenomenological model relates \( \sigma(t) \) to the absorption cross section:
\begin{equation}
\sigma(t) \sim \int db \, b \, \left[1 - |\mathcal{A}(s,b)|^2 \right],
\end{equation}
which quantifies the total probability of loss of the probe to the black brane.

\subsection{Effective Absorption Model and Comparison Table}

We now present a comparison between entropy suppression and absorption probability in AdS:

\begin{table}[h!]
\centering
\caption{Geometric interpretation of entanglement suppression via scattering.}
\label{tab:scattering_vs_entropy}
\renewcommand{\arraystretch}{1.3}
\begin{tabular}{|p{4.2cm}|p{5.2cm}|p{5.2cm}|}
\hline
\textbf{Quantity} & \textbf{Field Theory} & \textbf{Bulk Gravity} \\
\hline
Operator quench & Local excitation & Infalling particle \\
\hline
Mixed state & Thermal background & AdS black hole \\
\hline
Suppression onset & \( t \sim t_\ast \) & Near-horizon interaction \\
\hline
Entropy loss & \( \sigma(t) \) & Absorption cross section \\
\hline
Correlation decay & \( \Delta S_A(t) \to \text{const.} \) & \( |\mathcal{A}(s,b)|^2 < 1 \) \\
\hline
Time delay & Geodesic stretching & Phase shift \( \delta(b) \) \\
\hline
\end{tabular}
\end{table}

The suppression of entanglement entropy in holographic field theories can be viewed as a shadow of gravitational scattering in the dual geometry. The loss of quantum information observed in von Neumann entropy or negativity is mirrored by absorption of bulk probes by black hole horizons. This scattering picture complements the HRT prescription and enriches our understanding of non-equilibrium quantum gravity. Further work may involve computing exact scattering amplitudes in time-dependent backgrounds and relating them directly to time-resolved entanglement spectra.

\section{Conclusion}

These five directions collectively aim to establish entanglement suppression as a universal dynamical phenomenon with deep ties to quantum gravity, chaos, and information theory. Whether approached through holography, conformal field theory, or quantum simulations, each line of inquiry promises both conceptual advances and calculable predictions.

The recent work of Doi and Takayanagi \cite{doi2025entanglement} serves as a foundational reference point, demonstrating that entanglement entropy in holographic CFTs can be significantly suppressed under a double-quench setup involving a localized excitation and a thermal background. Their analysis, combining insights from AdS/CFT, quantum information, and bulk gravitational dynamics, offers a compelling case for interpreting such suppression as a boundary manifestation of bulk scattering and absorption.

By generalizing their framework to higher dimensions, incorporating quantum chaos diagnostics like OTOCs, analyzing integrable models, and extending to mixed-state entanglement measures such as negativity, this roadmap builds upon and significantly broadens the scope of their results. The geometric re-interpretation of suppression as linked to scattering cross sections not only complements their findings but opens new paths to connect entanglement dynamics with semi-classical and quantum gravitational phenomena.

Ultimately, these investigations aim to deepen our understanding of the flow of quantum information in time-dependent spacetimes. They highlight how holography can serve as a bridge between entanglement-based diagnostics and gravitational observables, reinforcing the idea that spacetime geometry and quantum entanglement are intimately intertwined. Future work may involve experimental analogues in quantum simulators or tabletop systems to test predictions of entanglement suppression, making this theoretical frontier increasingly accessible and testable. 
These recent contributions \cite{lilani2025butterfly, lu2023order} underscore the growing interest in entanglement suppression and its connections to holography, chaos, and quantum information. They provide a timely foundation for extending the current framework to higher dimensions, experimental platforms, and refined diagnostics.
\subsection*{Future Directions}
Future research could explore entanglement suppression in higher-dimensional CFTs, investigate its manifestation in quantum simulation platforms, and develop experimental analogues using cold atoms or superconducting qubits. Additionally, refining the holographic prescription for entanglement negativity and its relation to multipartite scrambling remains a promising avenue.

\end{document}